\documentclass[a4paper,12pt,brazil]{article}
\pdfoutput=1
\usepackage[utf8]{inputenc}
\usepackage[brazil]{babel}
\usepackage[margin=2cm]{geometry}
\usepackage{amsmath,amssymb}
\usepackage{url}
\usepackage{graphicx}
\usepackage{tabularx}
\usepackage[numbers,sort&compress]{natbib}
\usepackage{algorithm}
\usepackage{algpseudocode}
\usepackage{indentfirst}
\usepackage{lastpage}
\usepackage{bm}
\usepackage{geometry}
\usepackage{tabularray}
\usepackage[table,xcdraw,pdftex,dvipsnames]{xcolor}
\usepackage[usestackEOL]{stackengine}
\usepackage{placeins}
\usepackage{textcomp}
\usepackage[np,autolanguage]{numprint}
\stackMath
\makeatletter
\renewcommand{\ALG@name}{Algoritmo}
\makeatother

\title{Algoritmos para Multiplicação Matricial} 
\author{Murilo Stellfeld de Oliveira Poloi\\Prof. Thiago de Oliveira Quinelato}

\newcommand{\N}{\ensuremath{\mathbb{N}}}

\begin{document}

\maketitle

\section{Introdução}
O objetivo deste artigo é estudar algoritmos que computam o produto entre duas matrizes, mais especificamente utilizando os métodos ingênuo, de {\it Strassen} e de {\it Strassen-Winograd}, que serão apresentados na Seção~\ref{sec:fundamentacao}.

Atualmente, os métodos citados não são os mais otimizados considerando a complexidade aritmética de seus algoritmos (vide Tabela \ref{tabela1}). No entanto, serão expostas modificações dos métodos de {\it Strassen} e {\it Strassen-Winograd} que conseguem reduzir sua alocação de memória e/ou tempo de execução. 

Os algoritmos do problema em estudo foram implementados utilizando a linguagem de programação {\it Julia}, na versão 1.9.1, com o auxílio dos pacotes {\it Pluto} ({\it notebooks}), {\it Plots} (visualização gráfica dos resultados) e {\it BenchmarkTools} (medição de alocação de memória e tempo de execução dos algoritmos). 

\begin{table}[!htb]
\centering
\begin{tabular}{|l|l|l|}
\hline
\textbf{Autoria}                 & \textbf{Ano}     & \textbf{$\omega \leq$} \\ \hline
\citet{Strassen_1969}                & 1969             & \np{2.8074}                 \\ \hline
\citet{Pan_1978}                     & 1978             & \np{2.796}                  \\ \hline
\citet{BINI1979234}                    & 1979             & \np{2.780}                  \\ \hline
\citet{schoenhage}               & 1981             & \np{2.522}                 \\ \hline
\citet{strassen86}                & 1986             & \np{2.479}                 \\ \hline
\citet{copperwin2} & 1990             & \np{2.3755}                  \\ \hline
\citet{Stothers2010OnTC}               & 2010             & \np{2.3737}                  \\ \hline
\citet{williams}                & 2013            & \np{2.3729}                \\ \hline
\citet{legall}                & 2014             & \np{2.3728639}                  \\ \hline
\citet{alman2021refined}                & 2020             & \np{2.3728596}                  \\ \hline
\citet{duan2022faster}                & 2022             & \np{2.37188}                  \\ \hline
\end{tabular}
\caption{Evolução da complexidade aritmética dos algoritmos de multiplicação de matrizes dada por $\mathcal{O}(n^{\omega})$, onde $n\in\N$ denota a ordem das matrizes. Fonte: \citet{algwiki}.}
\label{tabela1}
\end{table}

\section{Fundamentação Teórica}
\label{sec:fundamentacao}

Nesta seção serão apresentados os conceitos teóricos dos algoritmos mencionados na introdução, assim como formas de implementá-los de maneira básica, isto é, sem modificações de suas formas originais.

Para o caso do método ingênuo, será apresentado o Algoritmo \ref{alg1}, enquanto que para os métodos de {\it Strassen} e {\it Strassen-Winograd} serão apresentados os Algoritmos \ref{alg2} e \ref{alg3}.

Para os últimos dois, será utilizada a estratégia de {\it Divide and Conquer}, isto é, dividir o problema original em subproblemas de maneira recursiva.

\subsection{Método ingênuo}
A computação do produto entre matrizes de maneira ingênua é similar à maneira usual que se multiplicam duas matrizes: dadas matrizes $A_{m \times n}$ e $B_{n \times k}$, as entradas da matriz $C = A\times B$ são obtidas da seguinte forma:

$$
c_{ij} = \sum_{p = 1}^n a_{ip}b_{pj} \text{, com } i=1,\dots,m \text{ e } j = 1,\dots,k \text{.} 
$$

A implementação do método ingênuo pode ser observada no Algoritmo~\ref{alg1}.

\begin{algorithm}
\caption{Método ingênuo de calcular o produto entre as matrizes $A_{m \times n}$ e $B_{n \times k}$.}
\begin{algorithmic}
\Function{naive}{A, B}

\State $C \gets \text{zeros}\left( m,k \right)$
    \For{$i = 1:m$}
        \For{$j = 1:k $}
            \State $soma \gets 0$
            \For{$k = 1:n$}
                \State $soma \gets soma + A\left[i,k \right] \times B\left[ k, j \right]$
            \EndFor
            \State $C\left[ i, j \right] \gets soma$
        \EndFor
    \EndFor
    \State\Return $C$
\EndFunction
\end{algorithmic}
\label{alg1}
\end{algorithm}

A complexidade aritmética do algoritmo é dada por $\mathcal{O}(mnk)$, que pode ser generalizada para $\mathcal{O}(n^3)$ ao considerar que as matrizes de entrada são $A, B \in \mathbb{R}^{n \times n}$.

\subsection{Método de {\it Strassen}}

O método de \citet{Strassen_1969} baseia-se na pré-computação de produtos que são utilizados posteriormente para calcular as componentes da matriz resultante do produto das matrizes de entrada.
Para o caso de $A, B \in \mathbb{R}^{2\times 2}$, o método de {\it Strassen} pode ser descrito em três passos:

\begin{itemize}
    \item Primeiramente calculam-se sete produtos com os elementos das matrizes $A$ e $B$:

$$
P1 := (A_{11} + A_{22})(B_{11} + B_{22});
$$
$$
P2 := (A_{21} + A_{22})B_{11};
$$
$$
P3 := A_{11}(B_{12} - B_{22});    
$$
$$
P4 := A_{22}(-B_{11} + B_{21});
$$
$$
P5 := (A_{11} + A_{12})B_{22};   
$$
$$
P6 := (-A_{11} + A_{21})(B_{11} + B_{12});
$$
$$
P7 := (A_{12} - A_{22})(B_{21} + B_{22});
$$

\item Então, calculam-se as componentes da matriz resultante utilizando os sete produtos obtidos no passo anterior:
$$
C11 := P1 + P4 - P5 + P7;
$$
$$
C12:= P3 + P5 ;
$$
$$
C21 := P2 + P4; 
$$
$$
C22:= P1 + P3 - P2 + P6.
$$

\item Por fim, obtemos:
$$
A \times B = C = 
\begin{bmatrix} 
C11 & C12 \\
C21 & C22
\end{bmatrix}.
$$

\end{itemize}

Note que este algoritmo computa apenas multiplicações e adições. 

Para o caso de $A, B \in \mathbb{R}^{2\times 2}$, são calculadas exatamente sete multiplicações e dezoito adições, enquanto que para o caso geral, o método de {\it Strassen} possui uma complexidade aritmética de $\mathcal{O}(7n^{log_2 7} - 6n^2)$ operações, considerando o produto entre matrizes de ordem $2^k, k \in \mathbb{N}$, como descrito por \citet{Cenk_Hasan_2017}.

Para o caso geral do método de {\it Strassen}, o pseudocódigo é dado pelo Algoritmo~\ref{alg2}.

\begin{algorithm}
\caption{Método de {\it Strassen} para  matrizes $A, B$ de ordem $2^k \text{, } k \in \mathbb{N}$.}
\begin{algorithmic}
\Function{strassen}{A, B}

    \If {$A \in \mathbb{R}^{1 \times 1}$}
        \State\Return $A \times B$
    \EndIf
    \State {\bf divida as matrizes A e B em quadrantes}
    \State {\bf calcule} $P_1, P_2, \dots, P_7$ {\bf recursivamente}
    \State $C_{11} \gets P_1 + P_4 - P_5 + P_7 $
    \State $C_{12} \gets P_3 + P_5$
    \State $C_{21} \gets P_2 + P_4$
    \State $C_{22} \gets P_1 + P_3 - P_2 + P_6$
    \State $C \gets \begin{bmatrix}
        C_{11} & C_{12} \\ C_{21} & C_{22}
    \end{bmatrix}$
    \State \Return $C$
\EndFunction
\end{algorithmic}
\label{alg2}
\end{algorithm}

\subsection{Método de {\it Strassen-Winograd}}

Também chamado de forma de {\it Winograd} ou variante de {\it Winograd}, são algoritmos que possuem quinze adições e sete multiplicações, conforme discorrido pelos autores \citet{Cenk_Hasan_2017}. Serão apresentadas duas abordagens do método de {\it Strassen-Winograd}: 

\begin{itemize}
    \item a primeira foi descrita por \citet{Knuth1997}, que computa a matriz resultante com base na seguinte relação:
    $$
    \begin{bmatrix}
        a & b \\ c & d
    \end{bmatrix}
    \begin{bmatrix}
        A & C \\ B & D
    \end{bmatrix}
    =
    \begin{bmatrix}
        aA + bB & w + v + (a + b - c - d)D \\
        w + u + d(B + C - A - D) & w + u + v
    \end{bmatrix},
    $$

    em que $u = (c-a)(C-D)$, $v = (c+d)(C-A)$ e $w = aA + (c+d-a)(A+D-C)$;

    \item a segunda utiliza vinte e duas operações em bloco, como exposto por \citet{Boyer_Dumas_Pernet_Zhou_2009}, sendo elas:

\begin{itemize}
    \item oito adições:
    
    $$ 
    S1 := A21 + A22; 
    $$
    $$
    S2 := S1 - A11;
    $$
    $$
    S3 := A11 - A21; 
    $$
    $$ 
    T1 := B12 - B11; 
    $$
    $$
    T2 := B22 - T1; 
    $$
    $$
    T3 := B22 - B12; 
    $$ 
    $$
    S4 := A12 - S2;
    $$
    $$
    T4 := T2 - B21; 
    $$
    \item sete multiplicações recursivas:
    
    $$ 
    P1 := A11\times B11;
    $$ 
    $$
    P2 := A12 \times B21;
    $$
    $$
    P3 := S4\times B22;
    $$ 
    $$
    P4 := A22 \times T4; 
    $$
    $$
    P5 := S1\times T1; 
    $$ 
    $$
    P6 := S2 \times T2; 
    $$
    $$ 
    P7 := S3 \times T3; 
    $$
    \item sete adições:
    
    $$ 
    U1 := P1 + P2; 
    $$
    $$
    U2 := P1 + P6;
    $$ 
    $$
    U3 := U2 + P7; 
    $$
    $$ 
    U4 := U2 + P5; 
    $$
    $$
    U5 := U4 + P3;
    $$
    $$
    U6 := U3 - P4; 
    $$
    $$ 
    U7 := U3 + P5;
    $$
    \item que por fim resultam na matriz $C$, dada por:
    $$
    A \times B = C = 
    \begin{bmatrix}
        U1 & U5 \\ U6 & U7
    \end{bmatrix}.
    $$
\end{itemize}
\end{itemize}

O pseudocódigo para a primeira forma apresentada do método de {\it Strassen-Winograd} é dado pelo Algoritmo~\ref{alg3}. 

\begin{algorithm}
\caption{Método de {\it Strassen-Winograd} para  matrizes $A, B$ de ordem $2^k, k \in \mathbb{N}$.}
\begin{algorithmic}
\Function{winograd}{A, B}
\If {$A \in \mathbb{R}^{1\times 1}$}
    \State\Return $A \times B$
\EndIf
\State {\bf divida as matrizes A e B em quadrantes}

\State {\bf calcule} $U, V \text{ e } W$ {\bf recursivamente}

\State {\bf calcule} $C_{11}, C_{12}, C_{21} \text{ e } C_{22}$ {\bf recursivamente}

\State $C \gets 
        \begin{bmatrix}
            C_{11} & C_{12} \\ C_{21} & C_{22}
        \end{bmatrix}$

    \State\Return $C$
\EndFunction
\end{algorithmic}
\label{alg3}
\end{algorithm}

\FloatBarrier

\section{Estratégias e resultados}

Todos os testes foram realizados com um processador Intel\textsuperscript{\tiny\textregistered} Core\textsuperscript{\texttrademark} i7-4790 CPU @ 3.60GHz e 16 GB de memória RAM. Os algoritmos utilizados estão na forma de {\it notebooks} do pacote {\it Pluto} e podem ser acessados junto das figuras geradas em:

\begin{center}
\url{https://github.com/murlopoloi/matrixmultiplication}.
\end{center}

\begin{figure}
            \centering
            \includegraphics[scale = 0.6]{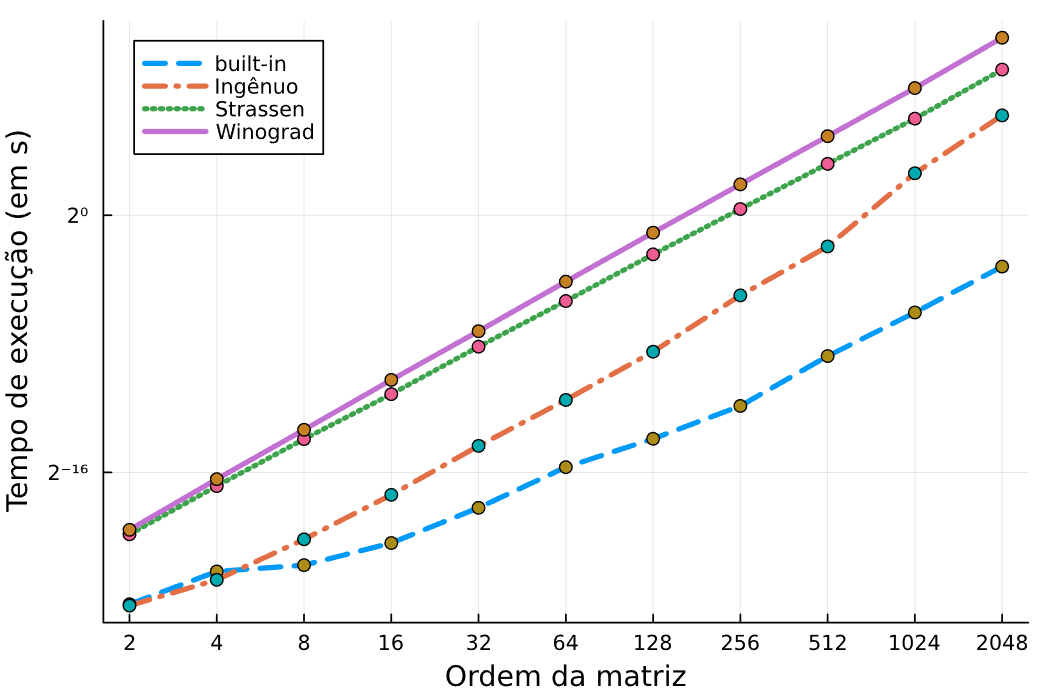}
            \caption{Tempo de execução dos algoritmos {\it built-in},  {\it Ingênuo}, {\it Strassen} e {\it Strassen-Winograd}.}
            \label{fig1}
        \end{figure}

\begin{figure}
             \centering
             \includegraphics[scale = 0.6]{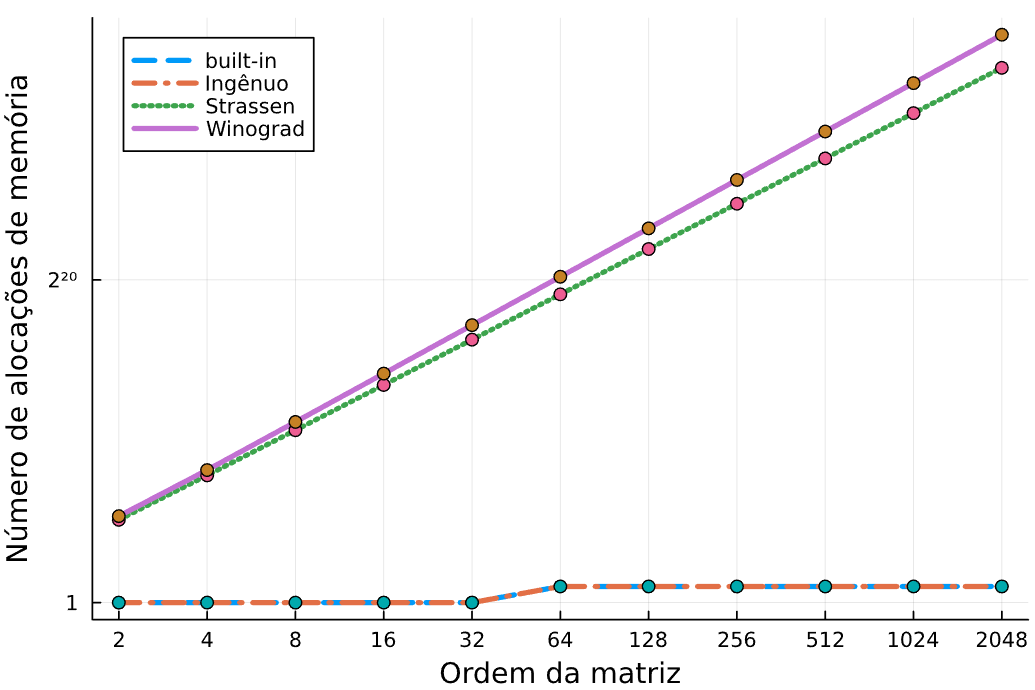}
             \caption{Alocação de memória dos algoritmos {\it built-in}, {\it Ingênuo}, {\it Strassen} e {\it Strassen-Winograd}.}
             \label{fig2}
        \end{figure}

Considerando tempo de execução e número de alocações de memória realizadas pelos Algoritmos \ref{alg1}, \ref{alg2} e \ref{alg3}, assim como o algoritmo {\it built-in} do {\it Julia} para computar o produto entre duas matrizes de ordem $2^k, k = 1, \dots, 11$, pode-se observar que ambos consumo de memória e tempo de execução dos métodos recursivos são maiores em comparação com o método ingênuo, vide Figuras~\ref{fig1}~e~\ref{fig2}.

A implementação do método de {\it Strassen-Winograd} foi a apresentada por \citet{Knuth1997} e possui uma chamada recursiva a mais em relação ao método de {\it Strassen}.


\subsection{Tratando matrizes não-quadradas}
Um dos problemas iniciais de métodos do tipo \textit{Divide and Conquer}, como {\it Strassen} e {\it Strassen-Winograd}, é a necessidade das matrizes de entrada serem da ordem $2^k$, com $k \in \mathbb{N}$. Quando isso não ocorre, podemos recorrer a uma técnica conhecida como {\it padding}, que consiste em adicionar linhas e/ou colunas de zeros às matrizes até obter-se o tamanho adequado para executar os métodos.

A implementação da técnica de {\it padding} é relativamente simples e não será um dos focos do artigo. Existem algoritmos de {\it padding} mais eficientes e outras abordagens para computar o produto entre matrizes não-quadradas em algoritmos que utilizam a técnica \textit{Divide and Conquer}.

\subsection{Interrompendo a recursão mais cedo}
Alternativamente, também é possível trocar o bloco 
\begin{center}
    \begin{minipage}{4.5cm}
        \begin{algorithmic}
        \If{$A, B \in \mathbb{R}^{1 \times 1}$} 
        \State\Return A $\times$ B
        \EndIf
        \end{algorithmic}        
    \end{minipage}
\end{center}
dos métodos de {\it Strassen} e de {\it Strassen-Winograd} a fim de evitar que a recursão chegue até o caso em que as matrizes possuam um só elemento, isto é, $A,B \in \mathbb{R}^{1 \times 1}$.

No caso da forma de {\it Winograd}, por exemplo, pode-se trocar o bloco acima por bloco(s) da forma
\begin{center}
    \begin{minipage}{5.5cm}
        \begin{algorithmic}
        \If{$A, B \in \mathbb{R}^{8 \times 8}$}
            \State\Return win8x8(A, B)
        \ElsIf{$A, B \in \mathbb{R}^{4 \times 4}$}
            \State\Return win4x4(A, B)
        \ElsIf{$A, B \in \mathbb{R}^{2 \times 2}$}
            \State\Return win2x2(A, B)
        \ElsIf{$A, B \in \mathbb{R}^{1 \times 1}$}
            \State\Return A $\times$ B
        \EndIf
        \end{algorithmic}   
    \end{minipage}
\end{center}
Neste bloco, \texttt{win2x2} é a função que calcula o método de {\it Strassen-Winograd} para matrizes em $\mathbb{R}^{2\times 2}$, \texttt{win4x4} para matrizes em $\mathbb{R}^{4\times 4}$ e assim por diante. Essa modificação será chamada de \textit{Strassen Mod} ou \textit{Winograd Mod}.

Naturalmente é possível parar a recursão para matrizes maiores. A mesma ideia pode ser replicada para o Algoritmo~\ref{alg2}, que representa o método de {\it Strassen}.

Essa estratégia afeta positivamente o tempo de execução e alocação de memória em comparação com os resultados observados nas Figuras \ref{fig1} e \ref{fig2}, visto que a modificação gera menos subproblemas a serem computados pelos métodos.

\begin{figure}[!htb]
            \centering
            \includegraphics[scale = 0.6]{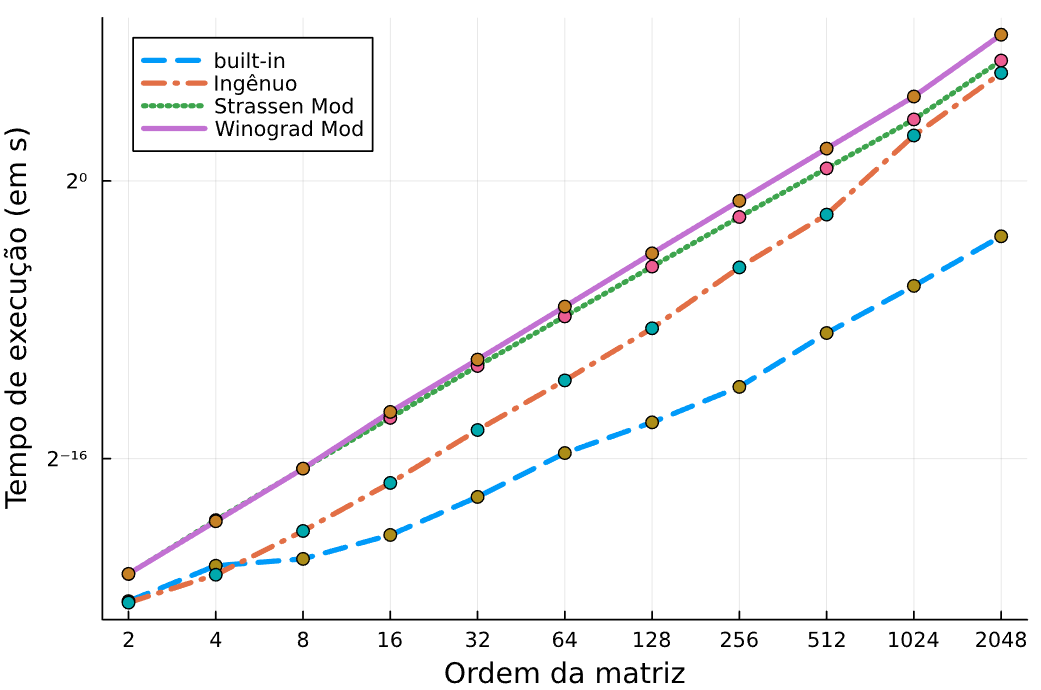}
            \caption{Tempo de execução dos algoritmos {\it built-in},  {\it Ingênuo}, {\it Strassen Mod} e {\it Winograd Mod}.}
            \label{fig5}
        \end{figure}

\begin{figure}[!htb]
    \centering
    \includegraphics[scale = 0.6]{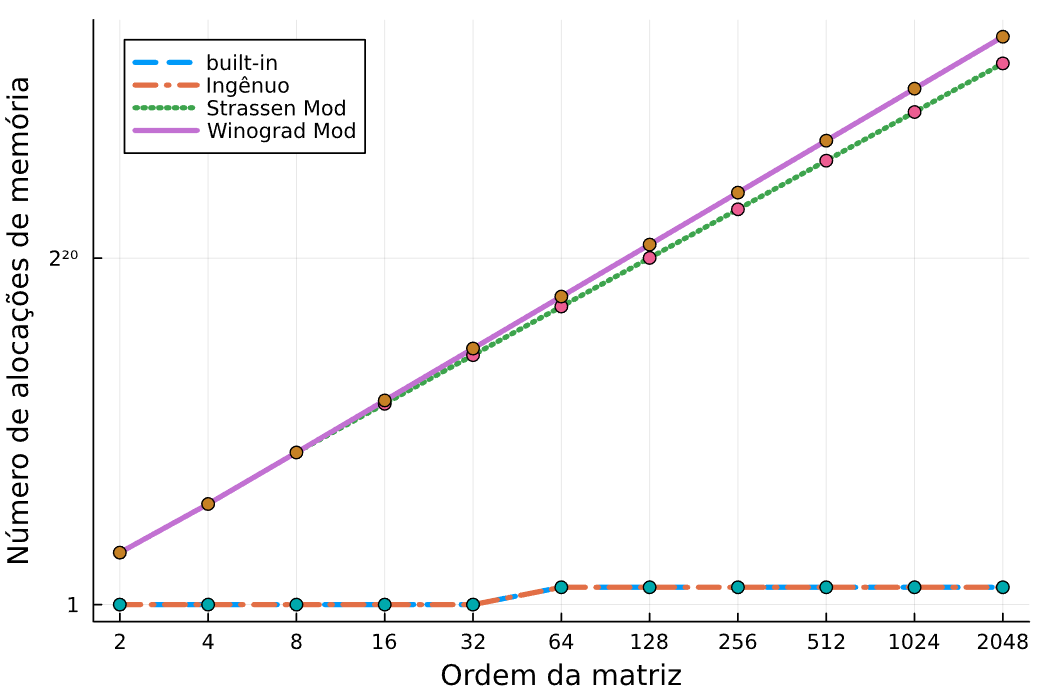}
    \caption{Alocação de memória dos algoritmos {\it built-in},  {\it Ingênuo}, {\it Strassen Mod} e {\it Winograd Mod}.}
    \label{fig6}
\end{figure}

O número de alocações de memória da Figura~\ref{fig6} é reduzido em relação à Figura~\ref{fig2}, mas ainda é de uma ordem maior que $2^{20}$ para matrizes de ordem maior ou igual a $128$. {\it Winograd Mod} continua tendo mais alocações considerando que também possui mais chamadas recursivas que {\it Strassen Mod}.

Ainda, é possível visualizar que o tempo de execução dos métodos na Figura~\ref{fig5} é reduzido em comparação com a Figura~\ref{fig1} e aproximam-se mais do tempo de execução do método ingênuo para matrizes de ordem maior ou igual a 1024, mas ainda não consomem menos tempo de execução que o mesmo.

\subsection{Otimização da alocação de memória}

Os autores \citet{Boyer_Dumas_Pernet_Zhou_2009} apresentam diversas abordagens a fim de reduzir o número de alocações de memória por parte do método de {\it Strassen-Winograd}. Foram implementadas duas delas:

\begin{itemize}
    \item {\it Two Temp Winograd}:
    são utilizadas duas variáveis temporárias para armazenar as informações necessárias na ordem de operações descrita na Tabela~\ref{tabela2};

    \begin{table}[!htb]
\centering
\begin{tabular}{|l|l|l|l|l|l|}
\hline
\textbf{\#:} & \textbf{Operação:}      & \textbf{Armazenado em:} & \textbf{\#:} & \textbf{Operação:}          & \textbf{Armazenado em:} \\ \hline
\textbf{1}   & $S_3 = A_{11} - A_{21}$ & X                    & \textbf{12}  & $P_1 = A_{11}B_{11}$ & X                    \\ \hline
\textbf{2}   & $T_3 = B_{22} - B_{12}$ & Y                    & \textbf{13}  & $U_2 = P_1 + P_6$           & $C_{12}$             \\ \hline
\textbf{3}   & $P_7 = S_3T_3$          & $C_{21}$             & \textbf{14}  & $U_3 = U_2 + P_7$           & $C_{21}$             \\ \hline
\textbf{4}   & $S_1 = A_{21} + A_{22}$ & X                    & \textbf{15}  & $U_4 = U_2 + P_5$           & $C_{12}$             \\ \hline
\textbf{5}   & $T_1 = B_{12} - B_{11}$ & Y                    & \textbf{16}  & $U_7 = U_3 + P_5$           & $C_{22}$             \\ \hline
\textbf{6}   & $P_5 = S_1T_1$          & $C_{22}$             & \textbf{17}  & $U_5 = U_4 + P_3$           & $C_{12}$             \\ \hline
\textbf{7}   & $S_2 = S_1 - A_{11}$    & X                    & \textbf{18}  & $T_4 = T_2 - B_{21}$        & Y                    \\ \hline
\textbf{8}   & $T_2 = B_{22} - T_1$    & Y                    & \textbf{19}  & $P_4 = A_{22}T_4$           & $C_{11}$             \\ \hline
\textbf{9}   & $P_6 = S_2T_2$          & $C_{12}$             & \textbf{20}  & $U_6 = U_3 - P_4$           & $C_{21}$             \\ \hline
\textbf{10}  & $S_4 = A_{12} - S_2$    & X                    & \textbf{21}  & $P_2 = A_{12}B_{21}$        & $C_{11}$             \\ \hline
\textbf{11}  & $P_3 = S_4B_{22}$       & $C_{11}$             & \textbf{22}  & $U_1 = P_1 + P_2$           & $C_{11}$             \\ \hline
\end{tabular}
\caption{Ordem de operações do algoritmo \textit{Two Temp Winograd} utilizando duas variáveis temporárias, $X$ e $Y$.}
\label{tabela2}
\end{table}

    \item {\it In-Place Winograd}:
    o resultado temporário das operações é armazenado diretamente nas matrizes de entrada $A$ e $B$ e na matriz de saída, $C$, na ordem de operações descrita na Tabela~\ref{tabela3}.
    \begin{table}[!htb]
\centering
\begin{tabular}{|l|l|l|l|l|l|}
\hline
\textbf{\#:} & \textbf{Operação:}      & \textbf{Armazenado em:} & \textbf{\#:} & \textbf{Operação:}   & \textbf{Armazenado em:} \\ \hline
\textbf{1}   & $S_3 = A_{11} - A_{21}$ & $C_{11}$             & \textbf{12}  & $S_4 = A_{12} - S_2$ & $A_{22}$             \\ \hline
\textbf{2}   & $S_1 = A_{21} + A_{22}$ & $A_{21}$             & \textbf{13}  & $P_6 = S_2T_2$       & $C_{22}$             \\ \hline
\textbf{3}   & $T_1 = B_{12} - B_{11}$ & $C_{22}$             & \textbf{14}  & $U_2 = P_1 + P_6$    & $C_{22}$             \\ \hline
\textbf{4}   & $T_3 = B_{22} - B_{12}$ & $B_{12}$             & \textbf{15}  & $P_2 = A_{12}B_{21}$ & $C_{12}$             \\ \hline
\textbf{5}   & $P_7 = S_3T_3$          & $C_{21}$             & \textbf{16}  & $U_1 = P_1 + P_2$    & $C_{11}$             \\ \hline
\textbf{6}   & $S_2 = S_1 - A_{11}$    & $C_{12}$             & \textbf{17}  & $U_4 = U_2 + P_5$    & $C_{12}$             \\ \hline
\textbf{7}   & $P_1 =  A_{11}B_{11}$   & $C_{11}$             & \textbf{18}  & $U_3 = U_2 + P_7$    & $C_{22}$             \\ \hline
\textbf{8}   & $T_2 = B_{22} - T_1$    & $B_{11}$             & \textbf{19}  & $U_6 = U_3 - P_4$    & $C_{21}$             \\ \hline
\textbf{9}   & $P_5 = S_1T_1$          & $A_{11}$             & \textbf{20}  & $U_7 = U_3 + P_5$    & $C_{22}$             \\ \hline
\textbf{10}  & $T_4 = T_2 - B_{21}$    & $C_{22}$             & \textbf{21}  & $P_3 = S_4B_{22}$    & $A_{12}$             \\ \hline
\textbf{11}  & $P_4 = A_{22}T_4$       & $A_{21}$             & \textbf{22}  & $U_5 = U_4 + P_3$    & $C_{12}$             \\ \hline
\end{tabular}
\caption{Ordem de operações do algoritmo \textit{In-Place Winograd}.}
\label{tabela3}
\end{table}

\end{itemize}

\begin{figure}[!htb]
    \centering
    \includegraphics[scale = 0.6]{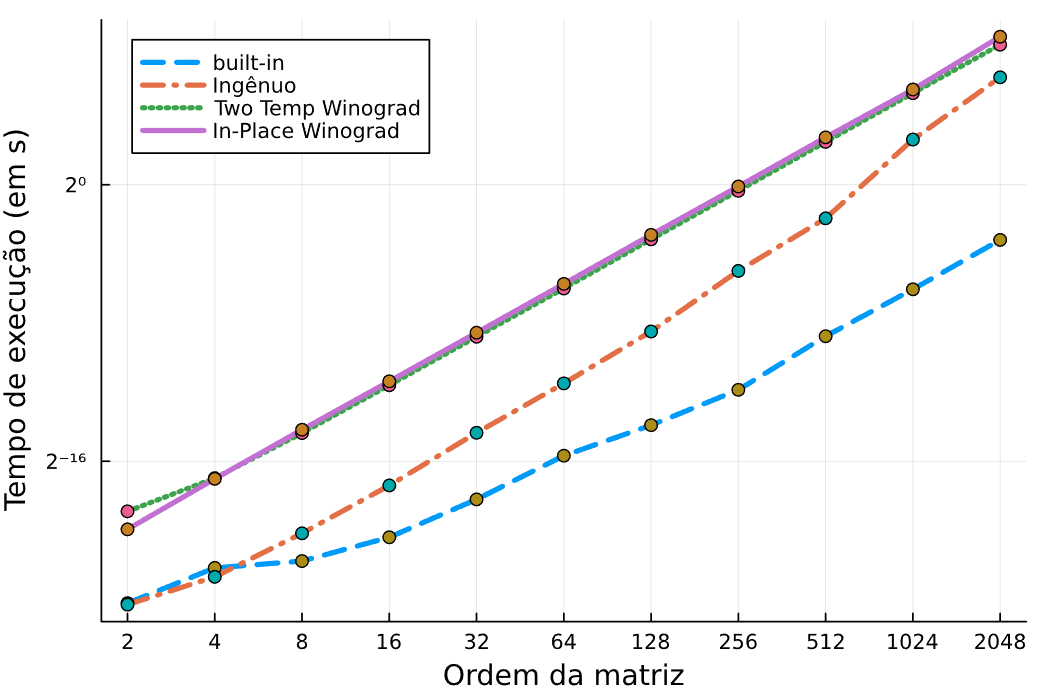}
    \caption{Tempo de execução dos algoritmos {\it built-in},  {\it Ingênuo}, {\it Two Temp Winograd} e {\it In-Place Winograd}.}
    \label{fig3}
\end{figure}

\begin{figure}[!htb]
    \centering
    \includegraphics[scale = 0.6]{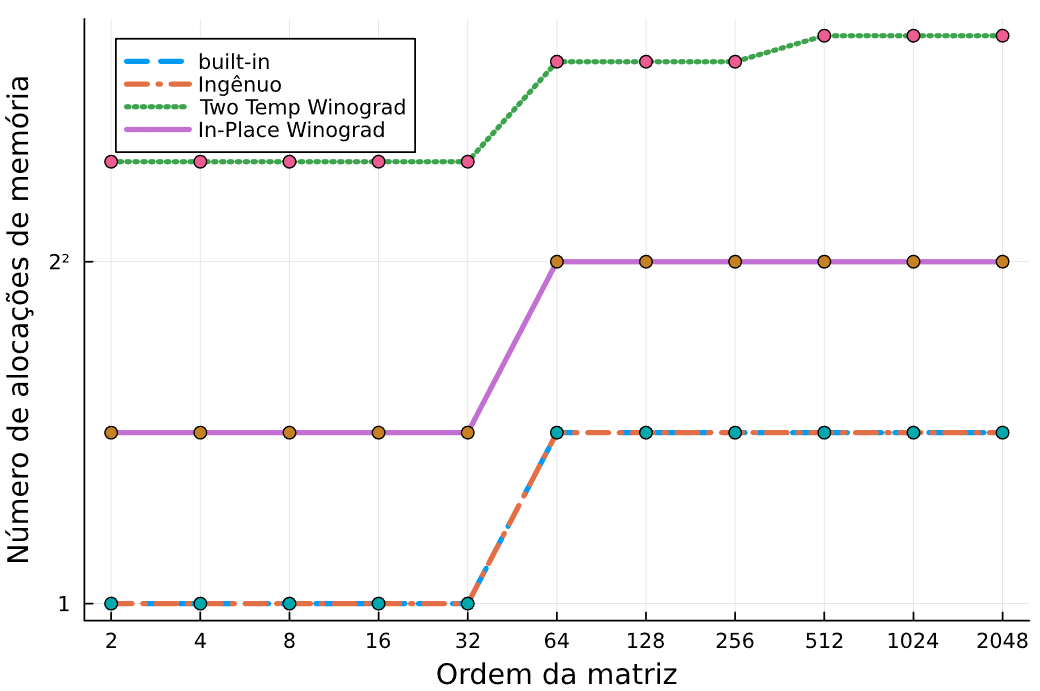}
    \caption{Alocação de memória dos algoritmos {\it built-in},  {\it Ingênuo}, {\it Two Temp Winograd} e {\it In-Place Winograd}.}
    \label{fig4}
\end{figure}

Os resultados de tempo de execução e alocação de memória das estratégias vistas acima podem ser observados nas Figuras~\ref{fig3}~e~\ref{fig4}.

O número de alocações dos algoritmos apresentados nas Tabelas~\ref{tabela2}~e~\ref{tabela3}, vide Figura~\ref{fig4}, é de fato reduzido em comparação com os resultados da Figura~\ref{fig2}. Em questão de tempo de execução, vide Figura~\ref{fig3}, ambas estratégias possuem tempos similares, mas ainda são mais lentas que o método ingênuo.

\subsection{Combinando o método ingênuo com {\it Strassen} e {\it Strassen-Winograd}}
Foi discorrido de maneira teórica e prática por \citet{Huss-Lederman_Jacobson_Tsao_Turnbull_Johnson_1996} que, sendo $n$ a ordem das matrizes de entrada, a partir de $n \leq 12$, torna-se mais viável utilizar o método ingênuo ao invés do método de {\it Strassen}. Esse valor é chamado de {\it cutoff criterion}.

Sejam $A_{m \times k}$ e $B_{k \times n}$. Os autores determinam o valor de $n$ para que  
$$
M(m,k,n) \leq 7M\left(\dfrac{m}{2}, \dfrac{k}{2}, \dfrac{n}{2}\right) + 4G\left(\dfrac{m}{2},\dfrac{k}{2}\right) + 4G\left(\dfrac{k}{2}, \dfrac{n}{2}\right) + 7G\left(\dfrac{m}{2}, \dfrac{n}{2}\right),
$$
em que $M(m,k,n)$ e $G(m,n)$, representam a complexidade aritmética, respectivamente, de se multiplicar as matrizes $A$ e $B$ pelo método ingênuo e de se adicionar ou subtrair matrizes de ordem $m \times n$. Do lado esquerdo da inequação temos a complexidade aritmética do método ingênuo, enquanto que do lado direito foi feita uma modelagem da complexidade aritmética do método de {\it Strassen}.

Utilizando contagem de operações, pode-se obter:
$$
mkn \leq 4(mk + kn + mn) \implies 1 \leq 4\left(\dfrac{1}{n} + \dfrac{1}{k} + \dfrac{1}{m}\right)\text{.}
$$

Esse resultado pode ser generalizado para matrizes $A, B \in \mathbb{R}^{n \times n}$ (daí $ n = k = m$), concluindo-se que é mais eficiente utilizar o método ingênuo ao chegar em matrizes de ordem $n \leq 12$ durante a recursão do método de {\it Strassen}. A mesma estratégia pode ser aplicada ao método de {\it Strassen-Winograd} visto que ambos possuem a mesma complexidade aritmética \citep{strwiki} e são do tipo {\it Divide and Conquer}.

Dessa forma, seja $2^k$ a ordem das matrizes em estudo. Podemos alterar o critério de parada da recursão por um bloco do tipo:
\begin{center}
\begin{minipage}{5cm}
\begin{algorithmic}
\If{$n\leq 12$}
    \State\Return \texttt{NAIVE}(A, B)
\EndIf
\end{algorithmic}
\end{minipage}
\end{center}

\begin{figure}
            \centering
            \includegraphics[scale = 0.6]{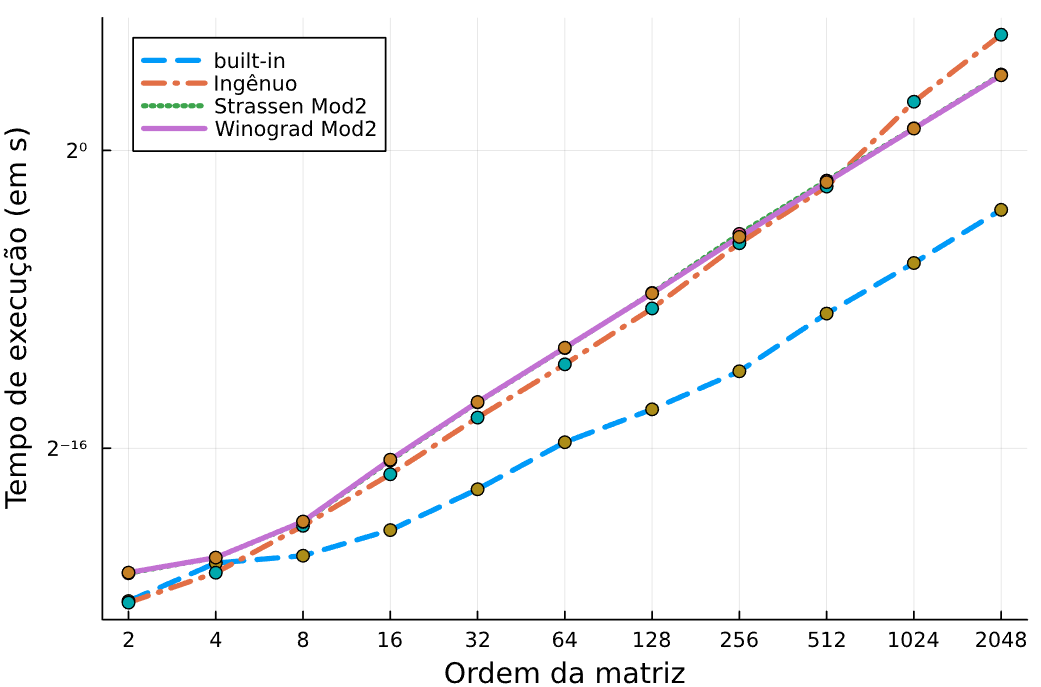}
            \caption{Tempo de execução dos algoritmos {\it built-in},  {\it Ingênuo}, {\it Strassen Mod2} e  {\it Winograd Mod2}.}
            \label{fig7}
        \end{figure}

\begin{figure}
    \centering
    \includegraphics[scale = 0.6]{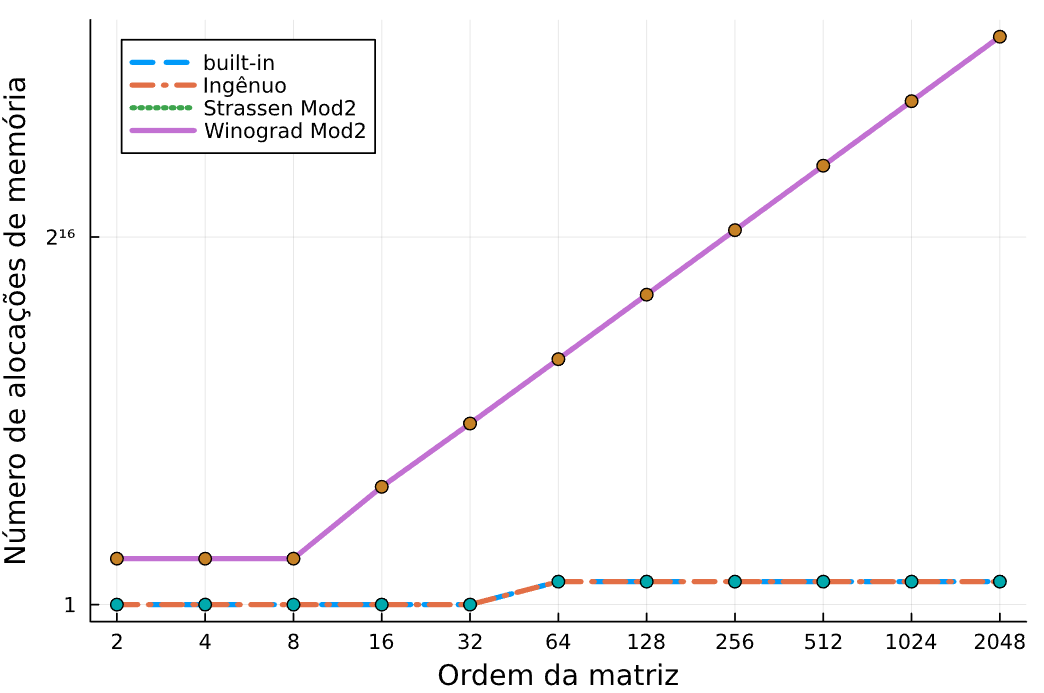}
    \caption{Alocação de memória dos algoritmos {\it built-in},  {\it Ingênuo}, {\it Strassen Mod2} e {\it Winograd Mod2}.}
    \label{fig8}
\end{figure}

Os algoritmos de \textit{Strassen} e \textit{Strassen-Winograd} modificados foram denotados por \textit{Strassen Mod2} e {\it Winograd Mod2}.
A estratégia de fato mostra-se adequada, conforme os resultados mostrados nas Figuras~\ref{fig7}~e~\ref{fig8}. Os resultados para os métodos \textit{Strassen Mod2} e {\it Winograd Mod2} são quase coincidentes.

Assim, obtêm-se dois algoritmos com tempo de execução menor em comparação com o método ingênuo para ordem $2^k$ das matrizes de entrada suficientemente grande. A modificação feita nesta seção também influenciou na redução de alocação de memória.

\subsection{Otimização de memória e $n\leq 12$}
Para verificar se é possível otimizar tanto alocação de memória quanto tempo de execução dos algoritmos utilizando as técnicas vistas anteriormente, foi aplicada a modificação da seção anterior nos algoritmos {\it Two Temp Winograd} e {\it In-Place Winograd} (modificação chamada, respectivamente, de  {\it Two Temp Winograd Mod} e {\it In-Place Winograd Mod}). 

\begin{figure}
            \centering
            \includegraphics[scale = 0.6]{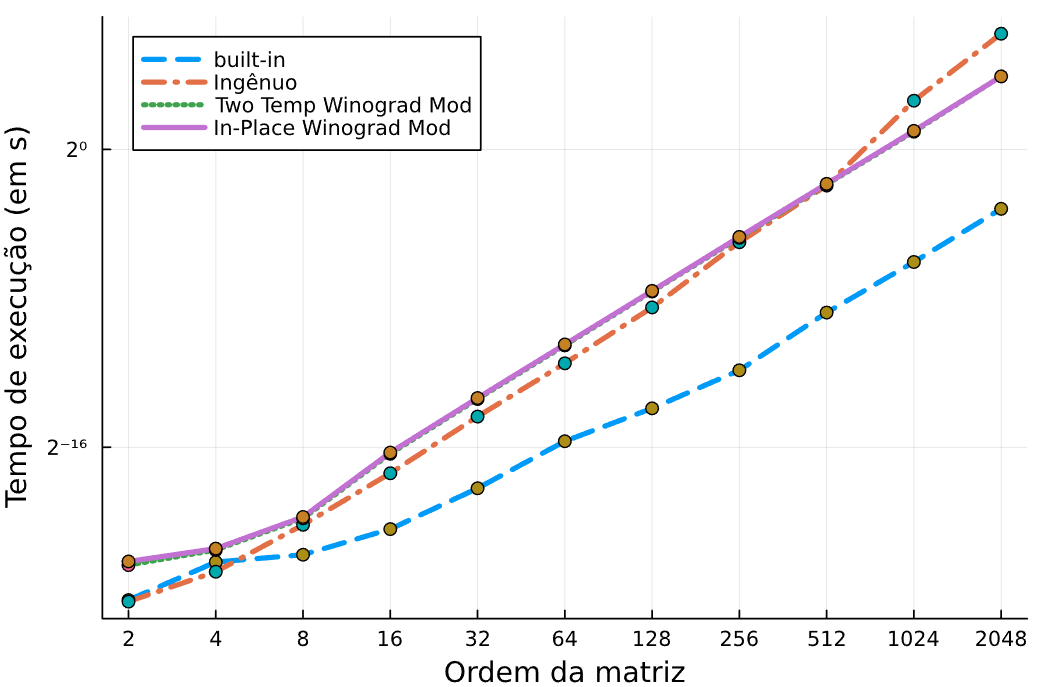}
            \caption{Tempo de execução dos algoritmos {\it built-in},  {\it Ingênuo}, {\it Two Temp Winograd Mod} e {\it In-Place Winograd Mod} .}
            \label{fig9}
        \end{figure}

\begin{figure}
    \centering
    \includegraphics[scale = 0.6]{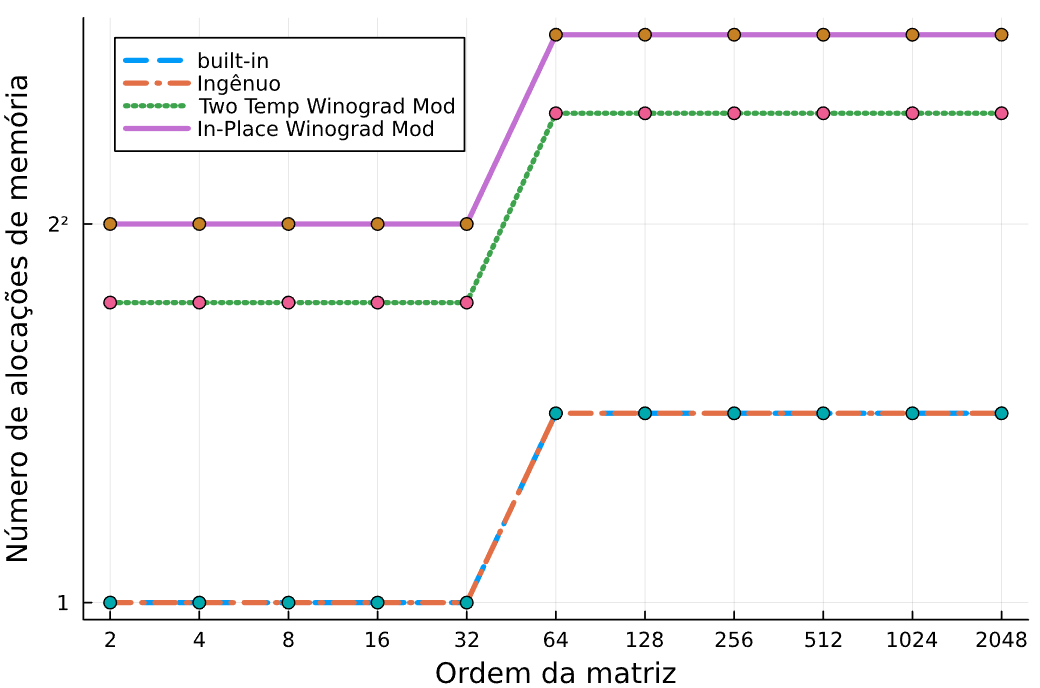}
    \caption{Alocação de memória dos algoritmos {\it built-in},  {\it Ingênuo}, {\it Two Temp Winograd Mod} e {\it In-Place Winograd Mod}.}
    \label{fig10} 
\end{figure}

Os resultados dessa modificação podem ser observados nas Figuras~\ref{fig9}~e~\ref{fig10}. Os métodos {\it Two Temp Winograd} e {\it In-Place} Winograd são quase coincidentes na Figura~\ref{fig9}.

A otimização em relação ao tempo de execução dos algoritmos manteve-se similar à mudança apresentada na seção anterior; o número de alocações de memória foi similar em relação aos algoritmos {\it Two Temp Winograd} e {\it In-Place Winograd}.

\FloatBarrier

\section{Conclusão}

Como mencionado na introdução, os métodos estudados são obsoletos em termos de complexidade aritmética, porém, são de simples estudo e implementação.

Os métodos de {\it Strassen} e de {\it Strassen-Winograd} sem modificações possuem desempenhos ruins em comparação com suas versões modificadas. Ademais, das estratégias exploradas, foi possível reduzir simultaneamente o tempo de execução e alocação de memória dos algoritmos de modo que:
\begin{itemize}
    \item para matrizes suficientemente grandes o tempo de execução fosse menor que o do método ingênuo;
    \item a alocação de memória se mantivesse similar ou igual à dos métodos {\it built-in} e ingênuo.
\end{itemize}

A otimização simultânea foi possível a partir da combinação de estratégias individuais de otimização de memória, com os métodos {\it Two Temp Winograd} e {\it In-Place Winograd}, e de tempo de execução, com a  estratégia apresentada por \citet{Huss-Lederman_Jacobson_Tsao_Turnbull_Johnson_1996}.

\bibliographystyle{sbpo}
\bibliography{ref} 

\end{document}